\documentclass[sigconf]{acmart}

\usepackage{booktabs} % For formal tables
\usepackage{hyphenat}
\usepackage{subfigure}
\usepackage{xspace}
\usepackage{listings}

\usepackage[hide]{chato-notes}
\newcommand{\cpt}[1]{\textsc{\MakeLowercase{#1}}}

\newcommand{\ie}{\textit{i.e.}\xspace}
\newcommand{\eg}{\textit{e.g.}\xspace}
\newcommand{\cf}{\textit{cf.}\xspace}

\newcommand{\vs}{\textit{vs.}\xspace}
\newcommand{\etc}{\textit{etc.}\xspace}
\newcommand{\viz}{\textit{viz.}\xspace}

\newcommand{\hide}[1]{}

\newcommand{\xhdr}[1]{\vspace{1.7mm}\noindent{{\bf #1.}}}

\newcommand{\Secref}[1]{Sec.~\ref{#1}}

\newcommand{\Figref}[1]{Fig.~\ref{#1}}

\acmYear{2018} 
\setcopyright{none} % for the arXiv.org copy
\acmConference[SIGIR '18]{The 41st International ACM SIGIR Conference on Research \& Development in Information Retrieval}{July 8--12, 2018}{Ann Arbor, MI, USA}

\acmBooktitle{SIGIR '18: The 41st International ACM SIGIR Conference on Research \& Development in Information Retrieval, July 8--12, 2018, Ann Arbor, MI, USA}

\acmDOI{10.1145/3209978.3209984}
\acmISBN{978-1-4503-5657-2/18/07}
\acmPrice{}

\begin{document}
\title{Structuring Wikipedia Articles with Section Recommendations}

\author{Tiziano Piccardi}
%\authornote{}
%\orcid{}
\affiliation{%
  \institution{EPFL}
}
\email{tiziano.piccardi@epfl.ch}

\author{Michele Catasta}
\affiliation{%
  \institution{Stanford University}
}
\email{pirroh@cs.stanford.edu}

\author{Leila Zia} 
\affiliation{%
 \institution{Wikimedia Foundation}
}
\email{leila@wikimedia.org}

\author{Robert West}
\affiliation{%
  \institution{EPFL}
}
\email{robert.west@epfl.ch}

% \author{Tiziano Piccardi, Michele Catasta, Leila Zia, Robert West}
% %\authornote{}
% %\orcid{}
% \affiliation{%
%   \institution{EPFL}
% }
% \email{tiziano.piccardi@epfl.ch}

\begin{abstract}
Sections are the building blocks of Wikipedia articles. They enhance readability and can be used as a structured entry point for creating and expanding articles. Structuring a new or already existing Wikipedia article with sections is a hard task for humans, especially for newcomers or less experienced editors, as it requires significant knowledge about how a well-written article looks for each possible topic. Inspired by this need, the present paper defines the problem of section recommendation for Wikipedia articles and proposes several approaches for tackling it.
Our systems can help editors by recommending what sections to add to already existing or newly created Wikipedia articles. Our basic paradigm is to generate recommendations by sourcing sections from articles that are similar to the input article. We explore several ways of defining similarity for this purpose (based on topic modeling, collaborative filtering, and Wikipedia's category system). We use both automatic and human evaluation approaches for assessing the performance of our recommendation system, concluding that the category\hyp based approach works best, achieving precision@10 of about 80\% in the human evaluation.

%To further improve the performance of the recommendation system, we use matrix factorization techniques for collaborative filtering and learning-to-rank models for blending the information from multiple Wikipedia categories.

\end{abstract}


%\ccsdesc[500]{Computer systems organization~Embedded systems}

%\keywords{ACM proceedings, \LaTeX, text tagging}

\maketitle

\section{Introduction}
\label{sec:Introduction}

%\note{1.5p (incl.\ title, abstract, etc.)}

Wikipedia articles are organized in sections.
Sections improve the readability of articles and provide a natural pathway for editors to break down the task of expanding a Wikipedia article into smaller pieces. However, knowing what sections belong to what types of articles in Wikipedia is hard, especially for newcomers and less experienced users, as it requires having an overview of the broad ``landscape'' of Wikipedia article types and inferring what sections are common and appropriate within each type. 

%The above issue is compounded by the fact that
Despite the importance of sections,
a large fraction of Wikipedia articles does not have a satisfactory section structure yet:
less than 1\% of all the roughly 5 million English Wikipedia articles are considered to be of quality class ``good'' or better, and 37\% of all articles are stubs.%
\footnote{Stubs are articles considered too short to provide encyclopedic coverage of a subject.}
Finally, there are many inconsistencies in section usage, even within a given Wikipedia language; \eg, 80\% of the section titles created in English Wikipedia are used in only one article.

\begin{figure}
\includegraphics[height=1.8in, width=3in]{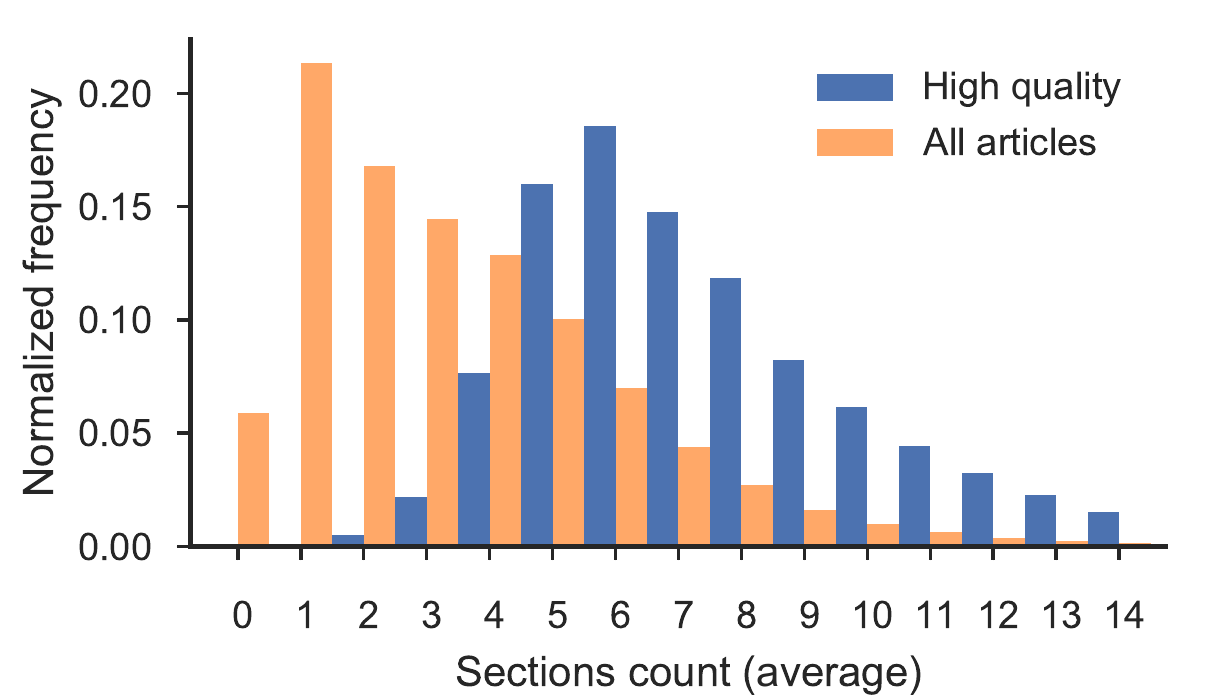}
\caption{Distribution of number of sections per article in English Wikipedia.
Good articles have more sections.
}
\label{fig:sections_distribution}
\end{figure}

Given Wikipedia's popularity and influence---with more than 500 million pageviews per day---, there is an urgent need to expand its existing articles across languages to improve their quality as well as their consistency. In other words, there is a need for a more systematic approach toward structuring Wikipedia articles by means of sections. 

%We also observe that the higher the quality of an article, the more the number of its sections. In fact, featured articles have twice as more sections as the the articles in stub or start categories. 

\begin{figure*}
\includegraphics[width=\textwidth]{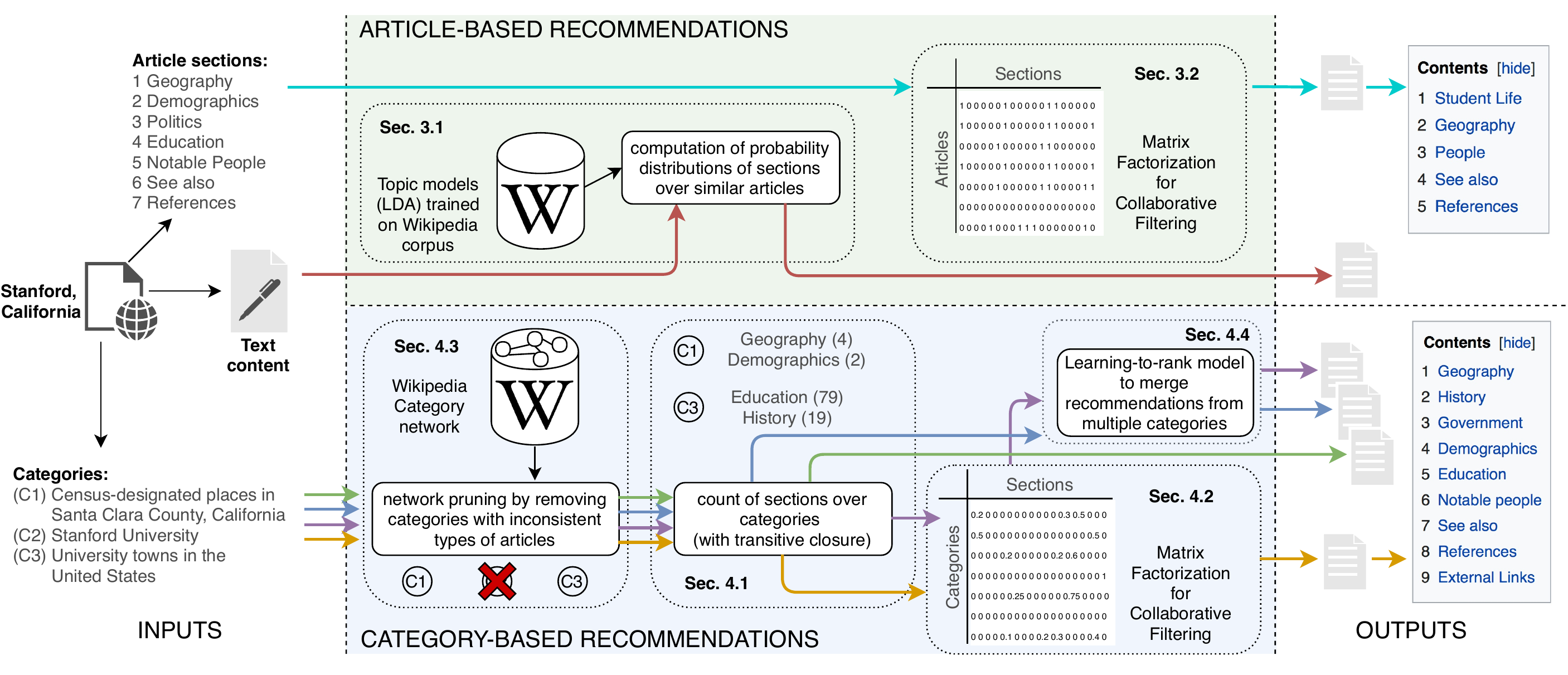}
\caption{
Overview of our systems for generating ranked lists of section recommendations (right) for a given input Wikipedia article (left).
This paper explores several approaches, each of which is represented as a path of arrows from left to right.
Different arrow colors designate different approaches.
At the broadest level, we consider two paradigms:
article\hyp based recommendation (top, shaded green) and category\hyp based recommendation (bottom, shaded blue).
Each component is labeled with a reference to the section of this paper that describes it.
%Note that the recommendation list in the upper right corner is purposefully of low quality, to illustrate the shortcomings of article\hyp based recommendation (\cf \Secref{sec:art-based eval: Using topic modeling}).
(Best viewed in color.)
}
\label{fig:overview}
\end{figure*}

\Figref{fig:sections_distribution} shows the distribution of the number of sections per article for all English Wikipedia articles, alongside the same distribution for the subset of articles considered to be of high quality, according to the Objective Revision Evaluation Service (ORES),%
\footnote{https://www.mediawiki.org/wiki/ORES}
a scoring system used to assess the quality of Wikipedia articles.
The plot shows that over one quarter of all articles have at most one section;
also, the number of sections is considerably lower when averaged over all articles (3.4), compared to the high\hyp quality subset (7.4).

The need for developing an approach to expand Wikipedia articles is acknowledged in the literature, where the majority of the methods developed focus on automatic expansion techniques. Algorithms are developed to propagate content across languages using the information in Wikipedia's information boxes \cite{Ta14}, to expand stubs by summarizing content from the Web \cite{Banerjee15} or from Wikipedia itself \cite{Banerjee15WP}, and to enrich articles using knowledge bases such as DBpedia \cite{Torres12}. However, these approaches are limited in their real\hyp life applicability to Wikipedia, for several reasons. First, the type of content they generate is limited (\eg, information boxes or short statements). Second, the accuracy of such approaches does not meet Wikipedia's high bar of content accuracy, which prevents such algorithms from being used in Wikipedia at scale. And third, these approaches are not editor\hyp centric, which is in fundamental contrast to how Wikipedia is built and run.

The desirability of an editor\hyp centric approach for empowering editors to expand Wikipedia articles, on the other hand, is acknowledged by experienced Wikipedia editors and developers through the creation of manually curated templates which are surfaced in raw format on Wikipedia howto pages as well as tools such as
% removing \textit{...} to make it consistent with other tool mentions (e.g., GapFinder).
Ma Commune
\cite{MaCommune} and
% removing \textit{...} to make it consistent with other tool mentions (e.g., GapFinder).
WikiDaheim
\cite{WikiDaheim}. These efforts aim to help newcomers to Wikipedia by providing recommendations on what sections to add to articles. The downside of these methods is that they do not scale and are very time-consuming to implement, as they rely on manually curated lists of sections per article type and for a given Wikipedia language. GapFinder \cite{Wulczyn16} and SuggestBot~\cite{Cosley07} are the only automatic editor-centric recommendation systems built and used in Wikipedia. GapFinder focuses on recommending what articles to create, while SuggestBot recommends articles through calls for specific actions, such as adding citations. None of the two systems, however, addresses the need for more in-depth recommendations on \textit{how} to expand an already existing article. 

In this paper, we take the first step for closing this gap in the literature by introducing and evaluating a series of editor-centric section recommendation methods. These methods differ in their source of signal (articles' topical content \vs\ Wikipedia's category network) as well as the technology used to model the recommendations (simple counting \vs\ collaborative filtering). We show that using Wikipedia's category network along with the proposed count-based approach provides the best recommendations, achieving precision@10 close to 80\% when evaluated by human editors.

\xhdr{Contributions}
Our main contributions are as follows.
\begin{itemize}
\item We introduce the problem of recommending sections for Wikipedia articles (\Secref{sec:System Overview}).
\item We devise multiple methods for addressing the problem (\Secref{sec:Article-based recommendation}--\ref{sec:Category-based recommendation}).
\item We test our methods thoroughly in both an automatic and a human evaluation (\Secref{sec:Evaluation dataset}--\ref{sec:Evaluation: Category-based recommendation}), finding that methods which leverage Wikipedia's category system clearly work best.
\end{itemize}

In the evaluation section, we present results based on the English version of Wikipedia, but our method is language\hyp independent, as it does not depend on any linguistic features.

\xhdr{Project repository}
We make code, data, and results available at \url{https://github.com/epfl-dlab/structuring-wikipedia-articles}.

% \begin{figure}
% %\includegraphics[height=1.8in, width=3in]{FIG/gini}
% \todo[Bob]{
% It would be good to have some proper ``Fig.~1'', something that motivates our entire work with one impressive number.
% Here, we could have something like: good articles tend to have $n$ secs per 1000 words; poor arts have much less (doesn't have to be this argument, can be anything)
% }
% \caption{
% }
% \label{fig:fig1}
% \end{figure}

\section{System Overview}
\label{sec:System Overview}

%\note{1p (incl.\ big diagram)}

This paper addresses the problem of recommending sections for Wikipedia articles, defined as follows:
given a Wikipedia article $A$ (which may be new or pre-existent), generate a ranking of sections to be added to $A$.

Before describing our solutions to this problem in detail, we give a high\hyp level overview of all the methods we explore.
We support our description with the illustration in \Figref{fig:overview}.

The \textbf{input} (left in \Figref{fig:overview}) to all our methods consists of a Wikipedia \textbf{article} $A$ that is to be expanded with additional sections.
$A$ is typically a stub that currently has very little content, and no, or only a few, sections.
The \textbf{output} (right in \Figref{fig:overview}) consists of a \textbf{ranked list} of recommended sections.

Our recommendations are intended to be screened by a human editor, who will decide which sections to incorporate into $A$, and who will ultimately write those sections. While there are no written rules to prevent machines from writing those sections directly in Wikipedia, in practice, such requirements are essentially imposed by Wikipedia's quality standards in a variety of its language editions.

%\footnote{This is vividly exemplified by a recent backlash inside the Wikipedia community against a research project aimed at automatically writing Wikipedia articles and inserting them into live Wikipedia \cite{bayer2016ai}.}

\Figref{fig:overview} exemplifies how a section ranking is produced for the input article $A=$ \cpt{Stanford, California} (the town, not the university).
Note that \Figref{fig:overview} illustrates all our methods in one single diagram, each method corresponding to one colored path of arrows.
At the broadest level, we explore two paradigms:
article\hyp based (top of \Figref{fig:overview}; \Secref{sec:Article-based recommendation}) and category\hyp based recommendation (bottom of \Figref{fig:overview}; \Secref{sec:Category-based recommendation}).

\textbf{Article\hyp based recommendation} (\Secref{sec:Article-based recommendation}) works directly with article features.
It leverages articles similar to the input article $A$ to suggest sections that are contained in many of them, but not yet in $A$ itself.
Similar articles are discovered in two ways:
The \textbf{topic\hyp modeling--based} method (\Secref{sec:Using topic modeling}) leverages articles that are similar to $A$ in terms of textual content.
A mere similarity in content does not, however, necessarily make another article a good source of sections;
\eg, the articles on \cpt{Stanford University} and \cpt{Leland Stanford} have much overlapping content, but one is about a school, the other about a person, which calls for rather different article structures and thus sections.
%as exemplified by the top right output ranking in \Figref{fig:overview}.
Hence, we also explore a \textbf{collaborative\hyp filtering--based} method (\Secref{sec:Using collaborative filtering}), which leverages articles that are similar to $A$ with respect to already\hyp present sections, rather than mere textual content. (Note that this method does not apply if $A$ has no sections yet.)
  
On Wikipedia, most articles are members of one or more so-called \textit{categories;} \eg \cpt{Stanford, California}, belongs to the category \cpt{University towns in the United States}, among others.
Our second broad paradigm, \textbf{category\hyp based recommendation} (\Secref{sec:Category-based recommendation}), makes use of this rich source of structured information by sourcing section recommendations for the input article $A$ from other articles belonging to the same categories as $A$.
In particular, for a given article $A$ and each category $C$ that $A$ belongs to, our \textbf{count-based} method (\Secref{sec:Using category--section counts}) computes a score for each section $S$, capturing what fraction of articles in $C$ contains $S$, and it then ranks sections by their scores.
This method yields one section ranking for each category $C$ that $A$ is a member of. If $A$ belongs to several categories, we \textbf{merge the rankings via learning\hyp to\hyp rank} (\Secref{sec:Combining recommendations from multiple categories}).

The number of possible recommendations is upper\hyp bounded by the total number of sections contained in articles in $C$, which may result in a small number of recommendations for very small categories.
We alleviate this problem by applying \textbf{collaborative filtering} (\Secref{sec:Generalizing via collaborative filtering}), which pools information between categories and allows us to make a large number of recommendations even for categories with only few member articles.

\begin{figure}
\includegraphics[width=\columnwidth]{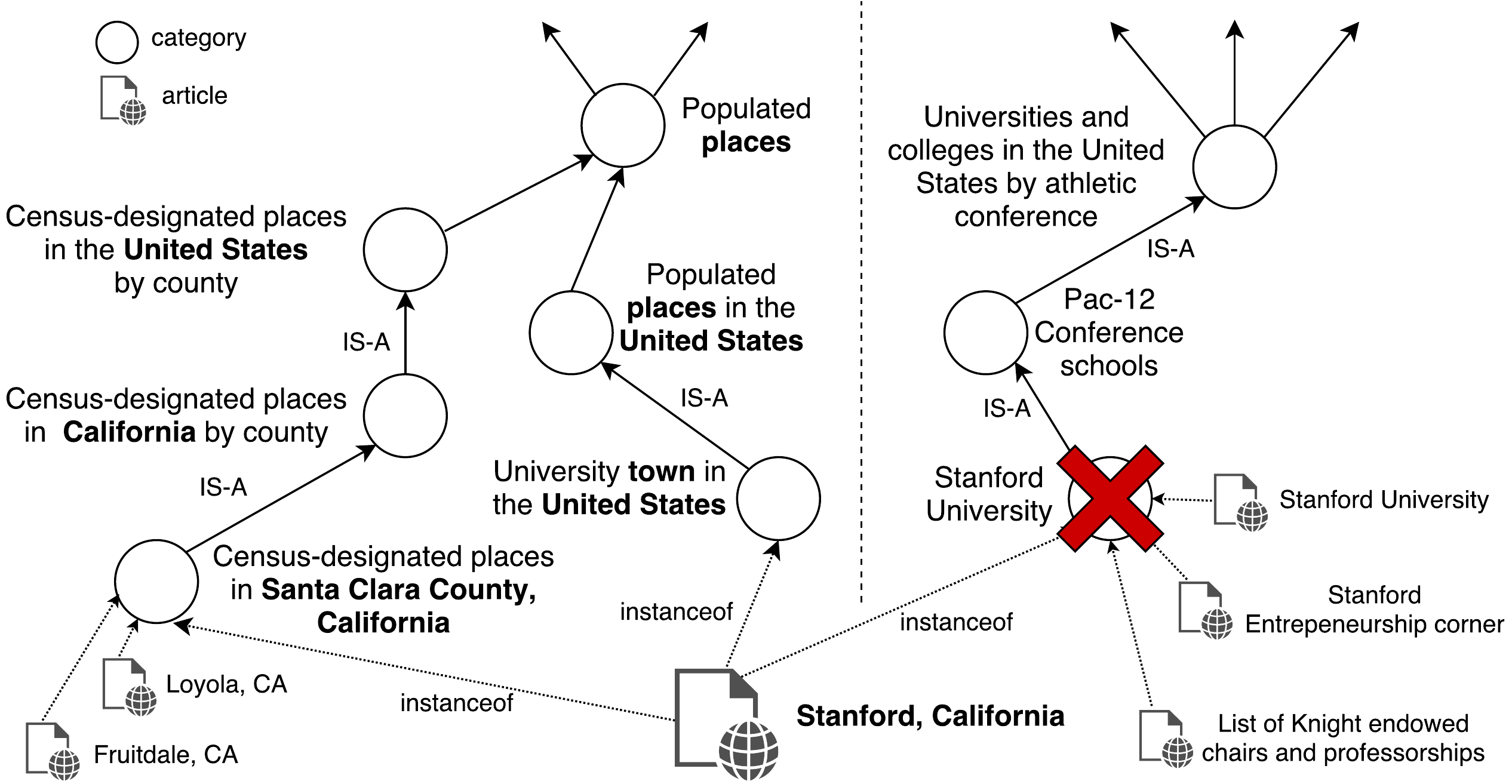}
\caption{Category network for the \cpt{Stanford, California} Wikipedia article ($A$). In this example, the two leftmost categories (assigned to $A$ via the \textit{instanceof} edges) are both subcategories of the base type \cpt{Populated places}. All the articles assigned to them (\cpt{Stanford, CA}, \cpt{Fruitdale, CA}, and \cpt{Loyola, CA}) have an \emph{ontological} relationship with their respective categories, \ie, they are instances of a populated place. We consider a category as \emph{pure} if the majority of its assigned articles respect the ontological relationship \emph{instanceof}. Conversely, the rightmost category \cpt{Stanford University} ($C$) is considered as impure, given the heterogeneous distribution of types of the articles assigned to $C$ (\ie, \cpt{List of Knight endowed chairs and professorships} is not an instance of \cpt{Stanford University}). Hence, our pruning algorithm removes $C$ from the network.
}
\label{fig:category_network}
\end{figure}

It is important to note that Wikipedia categories are organized in a network, with links representing the \textit{subcategory} relation.
Therefore, to render categories useful for our purpose, we need to reason about the transitive closure of this relation, rather than about single links;
\eg, \cpt{Stanford, California}, is not explicitly labeled as a populated place; this information is only implicit in the fact that the category \cpt{University towns in the United States} is connected to the category \cpt{Populated places} by a sequence of subcategory links (\Figref{fig:category_network}).
While straightforward in theory, working with the transitive closure is complicated in practice by the fact that Wikipedia's category network is noisy and ill\hyp conceived (\cf \Figref{fig:category_network} for an example), which gives rise to bad recommendations if not handled carefully.
To avoid this problem, we first \textbf{prune the category network} in a preprocessing step (\Secref{sec:Cleaning the category network}).

%\todo[Bob]{This caption is extremely confusing. Someone needs to rewrite this very carefully (it's very important for understanding the whole shebang...). The caption needs to explain why the one cat is removed, while the others aren't. It needs to be explained such that all the relevant nomenclature becomes clear (``ontological'', ``pure'').}

\section{Article-based recommendation}
\label{sec:Article-based recommendation}

%\note{Consider bringing the evaluations corresponding to the subsections to this section to here. As things stand now, if I read this section, I don't know what I should expect: will it work? will it not work? why am I being introduced to category-based recommendation next? I think it's confusing to say that we later show them why these two methods don't work.}

Intuitively, articles about similar concepts should have a similar section structure.
This intuition directly translates into a strategy for recommending sections for articles:
given an article $A$ for which sections are to be recommended, detect articles similar to $A$ and suggest sections that appear in those  similar articles but that do not appear in $A$ yet.

In order to turn this high\hyp level strategy into a concrete implementation, we need to define what we mean by ``similar articles'', and we need to specify a way of transforming the set of sections contained in those articles into a ranked list of recommendations.

In this section, we introduce two variants of the article\hyp based recommendation paradigm.
The first sources its recommendations from articles that are textually similar to $A$, as determined by a topic modeling method (\Secref{sec:Using topic modeling});
the second, from articles that are similar to $A$ with respect to sections that are already present, via collaborative filtering (\Secref{sec:Using collaborative filtering}).

We call these methods article\hyp based because they operate directly on $A$, instead of passing through the categories of which $A$ is a member, as done by the methods we shall introduce in \Secref{sec:Category-based recommendation}.

%- in the eval, we shall see that these methods have severe shortcomings, which is why we started exploring methods that pass through cats (we move to them in the sec after the present one)

\subsection{Using topic modeling}
\label{sec:Using topic modeling}

%To begin with, we leverage topical modeling of Wikipedia articles. The assumption is
Our topic\hyp modeling approach assumes
that articles containing similar topics should have a similar section structure. To this end, we build a topic model over all the articles in English Wikipedia via Latent Dirichlet Allocation (LDA) \cite{blei-et-al2003}. We use the LDA implementation included in the \textit{gensim} library \cite{rehurek_lrec} with the number of topics set to $K=200$.%
\footnote{Among the different values we tested in the range from 10 to 500, setting the parameter to 200 topics provided the best trade-off between accuracy and runtime performance. The quality of the recommendations generated by our baseline approach quickly plateaued for values above 200.}
In this model, an article $A$ can be thought of as a probability distribution over the $K$ topics.
%, represented as a $K$\hyp dimensional probability vector.
%Intuitively, the weights in the linear combination vector $W(A)$ represent the probability that the given article can be reconstructed by the terms of a certain topic. 

The main step of our topic\hyp modeling approach consists of generating a ranked list of section names per topic. To this end, we maintain $K$ dictionaries (one per topic), whose keys are all section names.
For each article $A$, we extract its list of sections, which is in turn used to update each dictionary:
for each topic $i$ and each section occurring in $A$, increment the respective entry of the $i$-th dictionary by $A$'s probability for topic $i$.
The output of this procedure is a set of $K$ ranked lists (one per topic), where each list contains all sections names for a specific topic in order of relevance.
To generate section recommendations for an article $A$, we
build a linear combination of all rankings, weighting the ranking corresponding to topic $i$ with $A$'s probability for topic $i$, thus effectively obtaining the most popular sections from the topics that best represent $A$.

This method combines the advantages of content-based and popularity-based recommender models.
On one hand, the content-based dimension of LDA allows us to mitigate the effect of ``cold-start'' articles (i.e., articles that cover novel content on Wikipedia can still be represented as a topic distribution by the topic model).
On the other hand, our approach inherently recommends sections that are popular among the entire corpus, given how we build the topic\hyp specific rankings. It is common knowledge \cite{steck2011item} that a popularity-based recommender system can be a hard-to-beat baseline, as popular items (in our specific case, sections) are often good candidates for recommendations.

\subsection{Using collaborative filtering}
\label{sec:Using collaborative filtering}

The above topic modeling approach uses only the textual content of articles and ignores information about the sections that already exist in the articles.
The approach we describe next does the opposite: it ignores textual content and focuses only on pre\hyp existing sections.
It aims to recommend sections from articles that are similar to the input article $A$ with respect to the sections that already exist in $A$.
(If $A$ has no sections yet, this method cannot be applied.)

This is a typical collaborative filtering (CF) setup.
Usually, CF is used to recommend items to users;
given a user to make recommendations for, it finds similar users and suggests items those users have liked.
In our setting, articles correspond to users, sections to items, and liking an item to the presence of a section.

We use alternating least squares \cite{Koren:2009:MFT:1608565.1608614}, a standard CF method based on matrix factorization.
We represent the data as a binary article--section matrix $M$, in which $M_{ij}=1$ if article $A_i$ contains section $S_j$, and $M_{ij}=0$ otherwise.
The matrix $M$ is then decomposed into two factor matrices $U$ and $V$ such that $M \approx UV^T = \tilde M$ and such that $U$ and $V$ have a low rank of at most $k$.
The rows of $U$ ($V$) represent articles (sections) in $k$ latent dimensions, so $\tilde M = UV^T$ captures the similarity of each article--section pair with respect to the latent dimensions and can thus be used to recommend new sections:
\eg, to suggest sections for article $A=A_i$, we sort the $i$-th row of $\tilde M$ in descending order and keep only the sections that are not already included in $A$.

% consider the $i$-th row of $\tilde M$.
% Each entry of this vector corresponds to a section, so we obtain a ranking of recommended sections by sorting this vector in descending order and keeping only the sections that are not already included in $A$.

\section{Category-based recommendation}
\label{sec:Category-based recommendation}

%\note{2p (all subsections + figures)}

Wikipedia articles are organized in a vast, user\hyp generated network of so-called \textit{categories}.
Ideally, links in this network would represent IS-A relationships and the network would be a taxonomical network, \ie, a tree\hyp structured, hierarchical grouping of articles into a coherent ontology by subject.
Unfortunately, this is not the case in practice, and before building recommendations using the category network, we need to clean it (\Figref{fig:category_network}).
This section will first explain how we can leverage the category network under the assumption that it represents a clean ontology (\Secref{sec:Using category--section counts}--\ref{sec:Generalizing via collaborative filtering});
then, in \Secref{sec:Cleaning the category network}, we will describe how the ill\hyp structured network can be preprocessed such that it more closely resembles a clean ontology by preserving only the likely IS-A relations in the graph.
Finally, in \Secref{sec:Combining recommendations from multiple categories}, we will describe how the actual recommendations are generated for a given input article $A$ by combing the relevant sections of different categories (if $A$ belongs to several categories).

%%%%%%%%%%%%%%%%%%%%%%%%%%%%%%%%%%%%%%%%%%%%%%%%%%

\subsection{Using category--section counts}
\label{sec:Using category--section counts}

%\note{0.25p}
Given a category $C$ as the entry point for generating recommendations, a simple and effective way to measure the relevance of a section $S$ is to count the number of times $S$ appears in all the articles in $C$.
Concretely, we proceed as follows:

\begin{enumerate}
\item As mentioned above, we assume for now that the category network provides a clean ontological structure, so we may obtain the set of all articles belonging to category $C$ via the \textit{transitive closure} of the subcategory relation:
an article $A$ is a member of category $C$ if it is either a direct member of $C$ or a member of a category in the transitive closure of $C$.
\item For each section title $S$, we count how many of $C$'s member articles contain a section titled $S$. Since category sizes---and thus section counts---may differ significantly, we normalize the counts by the number of articles in category $C$. The resulting scores may be interpreted as $P(S|C)$, \ie, the probability of observing each section $S$ in a randomly drawn article from category $C$.
\item The ranked list of recommendations is then given by sorting the set of all sections $S$ in decreasing order of $P(S|C)$.
\end{enumerate}

This procedure requires a clean, ontological category network, which is not given off the shelf (\Figref{fig:category_network}).
We address this issue in \Secref{sec:Cleaning the category network}, where we provide a method for cleaning the category network before passing it to the above\hyp described procedure.

%%%%%%%%%%%%%%%%%%%%%%%%%%%%%%%%%%%%%%%%%%%%%%%%%%

\subsection{Generalizing via collaborative filtering}
\label{sec:Generalizing via collaborative filtering}

%\note{0.25p}

The previous approach considers different categories as separate entities and can extract relevant sections only from the articles that are members of the respective category or of its subcategories. This is rather restrictive; \eg, while \cpt{Swiss scientists} and \cpt{American scientists} are semantically similar categories at the same level of the hierarchy and may thus share common sections, the counting\hyp based approach described in the previous section cannot pool information from these two disjoint subtrees.

Therefore, we are interested in grouping similar categories to learn about missing sections by transferring information between them. This is especially important for the smaller categories, such as \cpt{Nauruan scientists}, which may contain patterns similar to other categories but do not have enough content to contribute to the generation of recommendations on their own.

\begin{figure}
\includegraphics[height=1.8in, width=3in]{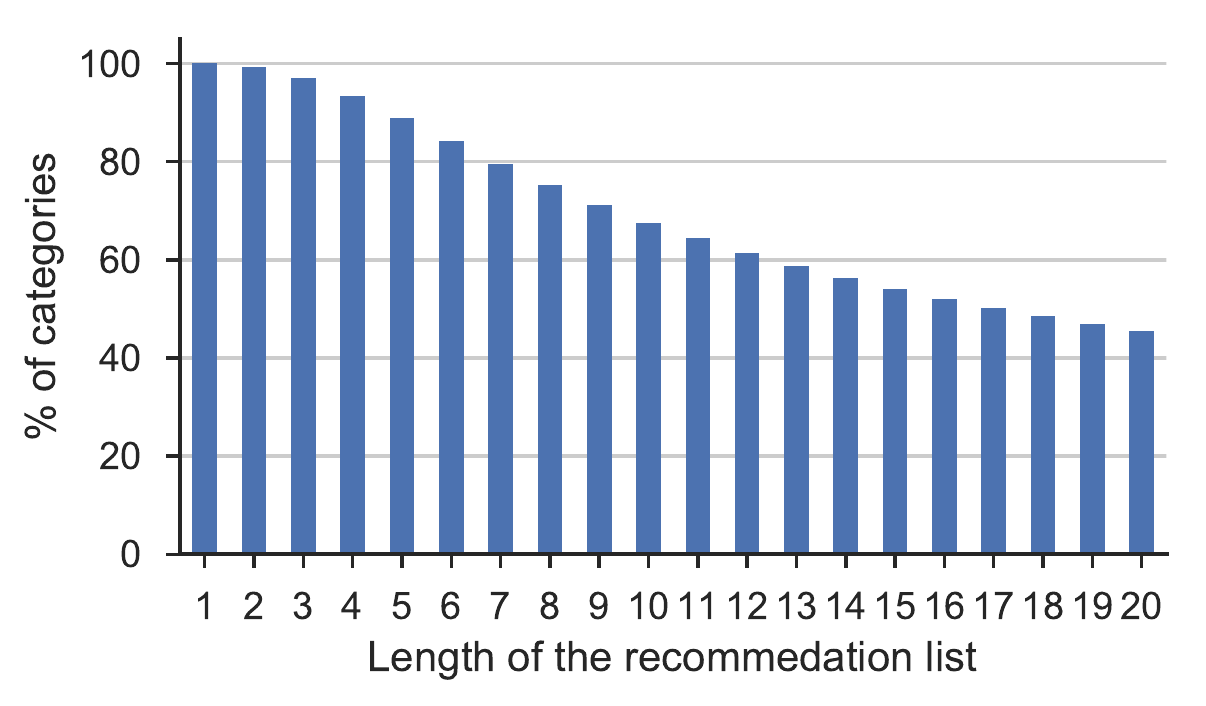}
\caption{Percentage of categories ($y$-axis) that can generate at least $x$ recommendations using the section\hyp count--based approach of \Secref{sec:Using category--section counts}.}
\label{fig:recommendations_length}
\end{figure}

To understand the impact of category size on the maximum number of recommendations we can generate, consider \Figref{fig:recommendations_length}, which shows, on the $y$-axis, the fraction of categories large enough to generate at least $x$ section recommendations when following the count\hyp based approach of \Secref{sec:Using category--section counts}. As shown by the plot, \eg, only 44\% of the categories have enough information to produce a recommendation list of 20 sections or more.

To overcome this limitation, we leverage the fact that collaborative filtering can recommend sections that never occurred in a certain category, by effectively extracting signals from similar categories. We apply the same matrix decomposition method that we used for generating recommendations from articles, rather than categories (\Secref{sec:Using collaborative filtering}). However, in the present setup, rows of the matrix $U$ represent categories (rather than articles) in latent space, and rows of the matrix $V$, sections (as before).
Hence, the matrix generated by the product $\tilde M = UV^T$ describes the relevance of a section for a category. We tested different ways to represent the scores in the matrix, and in the evaluation section we describe the setup that gives the best results.

%%%%%%%%%%%%%%%%%%%%%%%%%%%%%%%%%%%%%%%%%%%%%%%%%%

\subsection{Cleaning the category network}
\label{sec:Cleaning the category network}

%\note{0.75p}

As mentioned above, the category network is unfortunately rather noisy, and we cannot rely on the original network structure to extract a clean concept hierarchy. 
In Wikipedia, editors interpret and use the categories in different ways, and the resulting graph includes as links both subcategory relations and generic, unlabeled relations representing a variety of different ways in which two categories can be linked.
For instance, the category \cpt{Algorithms} is marked as a subcategory of \cpt{Applied mathematics}, although it is not the case that ``an algorithm is an applied mathematics''.
Indeed, \cpt{Applied mathematics} is not even a category in the sense of ``a collection of entities of the same type''; rather \cpt{Applied mathematics} is itself an entity. We refer to such ``categories'' as \textit{non\hyp ontological}.
What is needed is a method for cleaning the category network by removing non\hyp ontological categories and edges that do not encode subcategory relationships.

\xhdr{Prior approaches}
Extracting an ontological structure from the category network is a problem that has been addressed in the past with different methods and for different purposes, but it remains an open problem. 
For example, MultiWibi \cite{FLATI201666} extracts a taxonomy from Wikipedia's category network for multiple Wikipedia language editions, using English as a pivot language. Unfortunately, since the source code is not public, we could not run this method on a recent snapshot of Wikipedia. When experimenting with an older snapshot, we further noticed that MultiWibi removes too many categories from the network for our purposes.

%In the context of sections recommendations, and with the goal of developing a system usable with the current Wikipedia setup, we can relax the requirement of having a strict ontology. We are more interested in generating recommendations for as many categories as possible and designing a language\hyp independent approach that can preserve the topology of the categories network.

Another method \cite{DBLP:conf/coling/GuptaPKPP16}, proposed for English Wikipedia,
%is based on a combination of heuristics that help in cleaning the noise in the categories graph \cite{DBLP:conf/coling/GuptaPKPP16}. This approach
relies on language\hyp specific heuristics, which poses an undesirable limitation for us, as we strive to build a language\hyp independent system.

A third approach \cite{Boldi:2016:CWC:2872518.2891111} that we tried is based on the centrality scores of nodes in the category networks.
It imposes a total order on the set of all categories by ordering them in decreasing order of centrality and discarding all links that mark a lower\hyp centrality category as the parent of a higher\hyp centrality category.
While simple and intuitively appealing, we found this heuristic to not work well for our purposes in practice; \eg, it marks \cpt{Monkeys} as a subcategory of \cpt{Catarrhini} (a class of monkeys) because the latter is more central in the category network than the former, although the relationship should in fact be reversed.

%Their method can identify and remove user-generated noise and can be used as a cleaning strategy to run more complex data mining analysis on the category graph. The key idea is that a generic category should have more centrality in the network than its subcategories. The main shortcoming that makes this method not suitable for our case is that we observed this is not always the case. For example, the category \cpt{World War II} has a higher centrality than its parent \cpt{World Wars}.

\vspace{1.7mm}\noindent{{\bf Our approach.}\footnote{Code and data available at \url{https://github.com/epfl-dlab/WCNPruning}}}
We start from the basic insight that most of the problems of the category network are caused by non\hyp ontological categories, \ie, categories that do not represent collections of entities of the same type. As an example, consider the category (not the article) \cpt{Stanford University} of \Figref{fig:category_network}: Stanford University is itself an entity, not a collection of entities (no entity can be ``a Stanford University''), yet there is a Wikipedia category with its name; this category is in fact used like a tag, for marking articles and categories that are merely somehow related to Stanford University.
Treating this spurious, non\hyp ontological category the same way we treat proper, ontological categories has a polluting effect when building the transitive closure (\Secref{sec:Using category--section counts}) of the subcategory relation, as exemplified by \Figref{fig:category_network}:
\eg, as \cpt{Stanford Entrepreneurship Corner} is a member of the (spurious) category \cpt{Stanford University}, and the latter is a subcategory of \cpt{Pac-12 Conference schools}, \cpt{Stanford Entrepreneurship Corner} is erroneously classified as a member of \cpt{Pac-12 Conference schools}.
Such errors may pollute section counts for a vast number of categories as they percolate upward, and are amplified, in the category network.

Recognizing non\hyp ontological categories as the root of most problems in our setup, we aim at cleaning the category network by removing such categories, as well as all their incoming and outgoing edges.
Note that this may diffract the category network into several disjoint connected components. This is, however, not a problem in our specific use case, as we do not require a perfect, full\hyp coverage category network, but rather one that simply lets us compute reliable section counts for all categories.

The key operation in our algorithm is detecting non\hyp ontological categories.
In doing so, we build upon the simple intuition that a category that contains articles of more than one type (\eg, persons, places, and events) cannot be a proper, ontological category;
\eg, the aforementioned spurious category \cpt{Stanford University} has (among others) persons, organizations, and events as member articles and should therefore be removed.
Since the Wikipedia category network itself is not clean (which is the very reason for designing this algorithm), we of course cannot rely on it alone to determine whether a category is ontological.
Instead, we must use an external resource. We use DBPedia \cite{auer2007dbpedia}, but one might also use alternative sources, such as Freebase \cite{Bollacker:2008:FCC:1376616.1376746} or Wikidata \cite{Vrandecic:2014:WFC:2661061.2629489}.
Concretely, we extract from DBPedia, for each article, its top\hyp level type in the DBPedia type hierarchy (there are 55 top-level types).%
\footnote{The DBPedia top\hyp level type \cpt{Agent} is too general, so we break it up into its subtypes, \viz, \cpt{Person}, \cpt{Organization}, \cpt{Deity}, \etc}
For each category, we then construct a \emph{type histogram}, which summarizes the DBPedia types of the articles contained in the category, and model the homogeneity, or \textit{purity,} of the category as the Gini coefficient
of its type histogram. A low Gini coefficient means that a histogram distributes its probability mass more evenly over the 55 DBPedia types, which indicates an impure, non\hyp ontological category.

%This step allows to consider as pure the category \cpt{People from Lausanne} although the individual articles have different types: i.e. \emph{AthleticsPlayer}, \emph{MusicalArtist} or \emph{Engineer}.
%The result of this mapping is a list of 2.7M annotated articles that are used to compute the types frequencies distribution for each category.

%For this reason, we designed an algorithm to reduce the noise in the network by removing categories that contain inconsistent types of articles.

In order to prune the category network by removing impure categories, our method proceeds bottom up, starting from the sinks of the category network and propagating the set of their member articles to their respective parent categories, but only if the purity of the type histogram is above a predefined threshold. The propagated articles are then counted as part of the parent category and contribute to its type histogram.

Listing \ref{recursive_algorithm} shows pseudocode for a recursive implementation of this algorithm. We note, however, that, while the pseudocode captures the conceptual idea of our algorithm, we developed a more efficient implementation based on dynamic programming.

\begin{lstlisting}[label={recursive_algorithm}, caption={Category\hyp network pruning}, basicstyle=\small, language=Python, float=t]
def prune(current_node):
    type_hist = current_node.get_type_histogram()
    for c in current_node.children:
    	child_hist = prune(c)
        type_hist = type_hist + child_hist
    if purity(type_hist) > threshold:
        mark current_node as pure
        return type_hist
    else:
        remove current_node from category network
        return an empty histogram
\end{lstlisting}

%%%%%%%%%%%%%%%%%
% Don't need this figure; can explain everything in prose
%%%%%%%%%%%%%%%%%
% \begin{figure}
% \includegraphics[height=1.8in, width=3in]{FIG/gini}
% \caption{Precision/Recall at different Gini coefficient thresholds between 0.9 and 1}
% \label{fig:GiniThreashold}
% \end{figure}

%%%%%%%%%%%%%%%%%%%%%%%%%%%%%%%%%%%%%%%%%%%%%%%%%%

\subsection{Combining recommendations from multiple categories}
\label{sec:Combining recommendations from multiple categories}

One article can be part of multiple categories, so when making recommendations, we should take all of them into account by merging the recommendations produced by each category, in order to find a set of relevant sections for the article.
A simple way to merge the recommendations while promoting the most representative sections is to sum their scores when they are present in more than one list. As effective as this strategy could be, it does not take into account other interesting features of the categories, \eg, the number of its member articles, its type purity as captured by the Gini coefficient (\Secref{sec:Cleaning the category network}), \etc
Intuitively, such features capture the diversity of the recommendations generated by a category: the larger a category, and the lower its Gini coefficient, the more heterogeneous the recommended sections will generally be.

This problem setting is very similar to the one described in the vast body of literature on learning-to-rank (L2R) \cite{liu2009learning}. L2R was originally conceived to improve the performance of information retrieval systems, but its supervised machine learning models can also be applied to our scenario.

We develop a regression model with features based on the size of a category and its Gini coefficient. The features are generated by polynomial expansion (up to order 4), with the addition of logarithmic and exponential functions. After a round of feature selection and an optimization phase based on precision and recall (obtained on a validation set), we obtain a model capable of effectively merging the recommendations coming from multiple categories. We describe the results in \Secref{sec:Eval: Combining recommendations from multiple categories}.

\section{Evaluation dataset}
\label{sec:Evaluation dataset}

%\note{0.75p}

We test our methods using the English Wikipedia dump released on September~1, 2017, which encompasses 5.5M articles and 1.6M section occurrences. 
\Figref{fig:sections_distribution} shows the distribution of the number of sections per article in this dump and confirms that Wikipedia is in dire need of a strategy for supporting editors in expanding the content of articles, with 27\% of articles having none or only one section.
Parsing the dump, we also observed that 37\% of all pages are flagged as stubs.

% * <glederre@gmail.com> 2017-10-31T19:56:22.996Z:
% For the numbers above, you could do a reference to \ref{fig:sections_distribution}. 
% ^.

Taking a closer look at the data, we notice that the most frequent sections include titles that are generic and content\hyp independent, \eg, \cpt{See Also}, \cpt{References}, or \cpt{Links}. These sections add useful metadata to the article but do not contribute to the expansion of the content itself. If we discard a list of 14 section titles of this kind, over 54\% of all articles have none or only one section.

In the dataset, we can also observe the effect of a lack of indication provided to editors by the standard editing tool: more than 1.3M section titles appear only one single time, which is frequently because editors chose titles that are too specific.
%meaning that the editors added many inconsistent titles. 
For instance, the section \cpt{Monuments, images and cost} would be more consistent with the rest of Wikipedia if it were simply titled \cpt{Monuments}, and instead of \cpt{Examples of Optometers}, the more straightforward and more generic section \cpt{Examples} would have been better suited.

\xhdr{Dataset preprocessing} Before evaluating our methods, we clean the dataset to remove irrelevant sections that we do not want to include in our training phases. First, we remove all the unique sections, which do not represent a relevant signal for recommendation. Then, we remove a list of 14 sections (\cpt{References, External links, See also, Notes, Further reading, Bibliography, Sources, Footnotes, Notes and references, References and notes, External sources, Links, References and sources, External Links}), which we manually selected from among the most common titles. The rationale behind this is that these sections are generic enough to be recommended by default for nearly every article, so they should not occupy spots in the short and concise, context\hyp specific rankings we aim to provide. The evaluation results that we describe in the next section are based only on the remaining, context\hyp specific recommendations.
%\footnote{Note that this filtering does not affect the language-independence requirement and leaving them in the dataset would improve the performance: i.e. \cpt{References} is present in 76\% of the articles }
Moreover, all the articles flagged as stubs are removed to avoid training and evaluating the system on articles that have been labeled explicitly as inadequate.

%Given the original section distribution, the number of non\hyp empty articles shrank to 2,869,265, while the number of sections is 214,513.
After this cleaning step, 215K sections and 2.9M articles with at least one section remain.
This portion of the dataset represents a core set of articles that contain enough content from which to extract common patterns.
The distribution of section counts of \Figref{fig:sections_distribution} refers to this filtered subset.

\xhdr{Dataset split} We split the entire filtered article set into three parts: we use 80\% of the articles for training the different methods, 15\% for testing, and 5\% as a validation set for hyperparameter tuning.

\xhdr{Category network} To generate the pruned category network, we started from the category graph of the English edition released as a SQL dump in June 2017. From the original dataset composed of 1.6M nodes, we removed maintenance categories like \cpt{All articles needing additional references} or \cpt{Wikipedia pages referenced by the press} by keeping only categories in the subtree of the category \cpt{Main topic classification}. This step reduces the number of categories to 1.4M. Then, using a heuristic method for breaking cycles in the graph by isolating a so-called feedback arc set \cite{eades1993fast}, we removed 4,011 edges from a total of 3.9M.
After this step, the category network is a directed acyclic graph.

To select the best Gini\hyp coefficient threshold for pruning (\Secref{sec:Cleaning the category network}), we use a manually annotated dataset with 710 samples. We collected this data through a Web interface that showed an article $A$ and a category $C$ randomly selected from the
set of all of $A$'s ancestors.
%transitive closure of all the parents of the categories assigned to $A$.
For each sample, the annotators could see the complete article content and the question ``Is this a $C$?''. Feedback was provided via two buttons labeled ``Yes'' and ``No''.

This dataset contains a set of positive and negative examples of paths that should or should not, respectively, appear in the network. We pruned the graph with different Gini\hyp coefficient thresholds and computed precision and recall using this data as the testing set. 

Based on these results, we selected the break\hyp even point of precision and recall (0.71), fixing the Gini threshold to 0.966.
This threshold value drops from the category network the most impure 5.7\% of categories and removes many of the paths that would otherwise propagate the impurity to other nodes.

\xhdr{Performance metrics}
Given an article from the testing set, our goal is to reconstruct its set of sections as completely as possible, with as few top recommendations as possible.
For each article, we obtain its precision@$k$ as the fraction of the top $k$ recommended sections that are also contained in the testing article, and recall@$k$ as the fraction of sections from the testing article that also appear among the top $k$ recommendations.
Taking the average of all article\hyp specific precision@$k$ (recall@$k$) values yields the global precision@$k$ (recall@$k$), which we plot as a function of $k$, for $k=1,\dots,20$ (as 20 seems a reasonable number of recommended sections to show to a user in an editing tool).

\begin{figure*}
\centering
\subfigure[Topic modeling (\Secref{sec:Using topic modeling})]{
	\includegraphics[width=0.55\columnwidth]{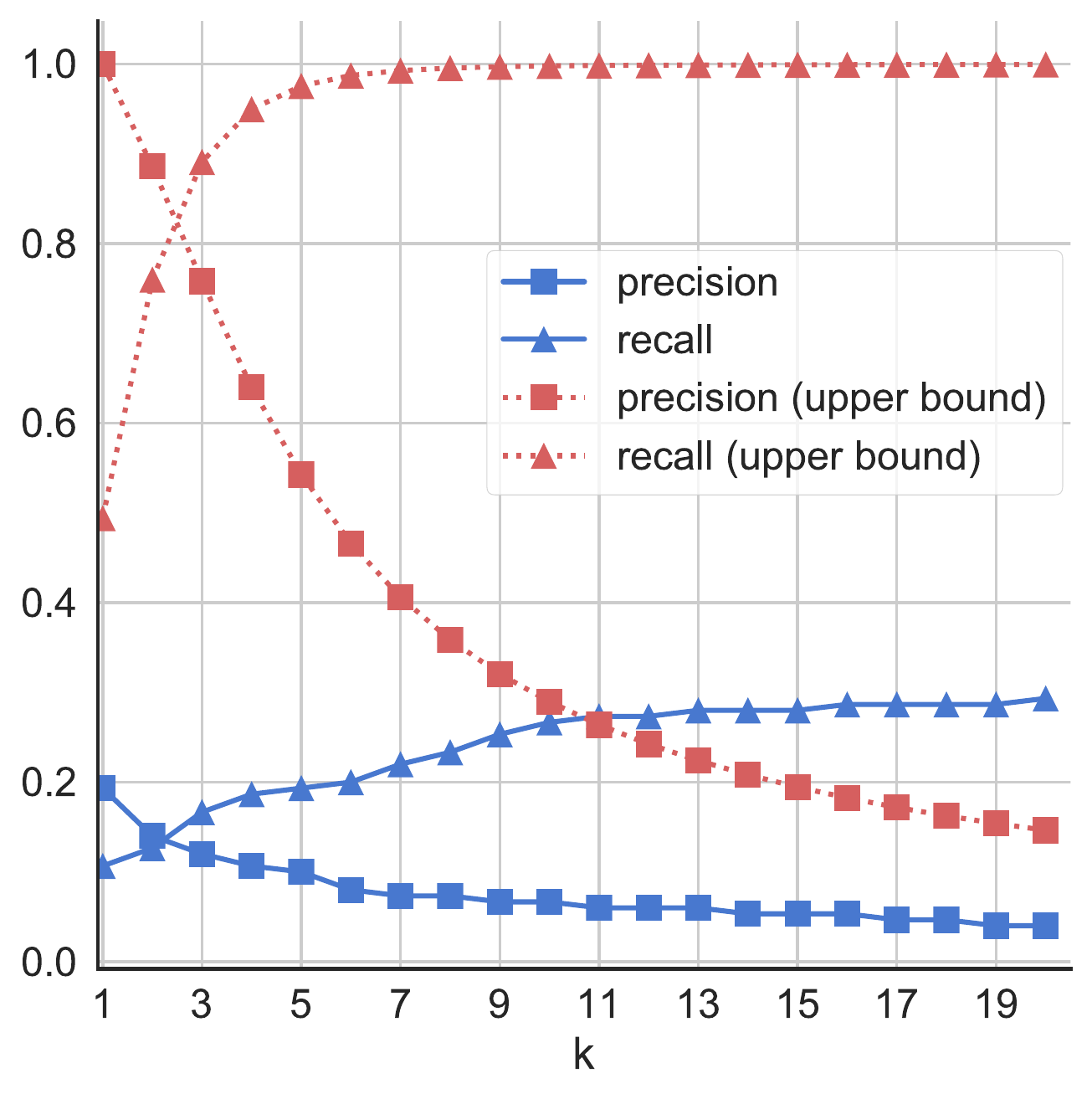}
    \label{fig:baseline_topic}
}%
\hfill
\subfigure[Category--section counts (\Secref{sec:Using category--section counts})]{
	\includegraphics[width=0.55\columnwidth]{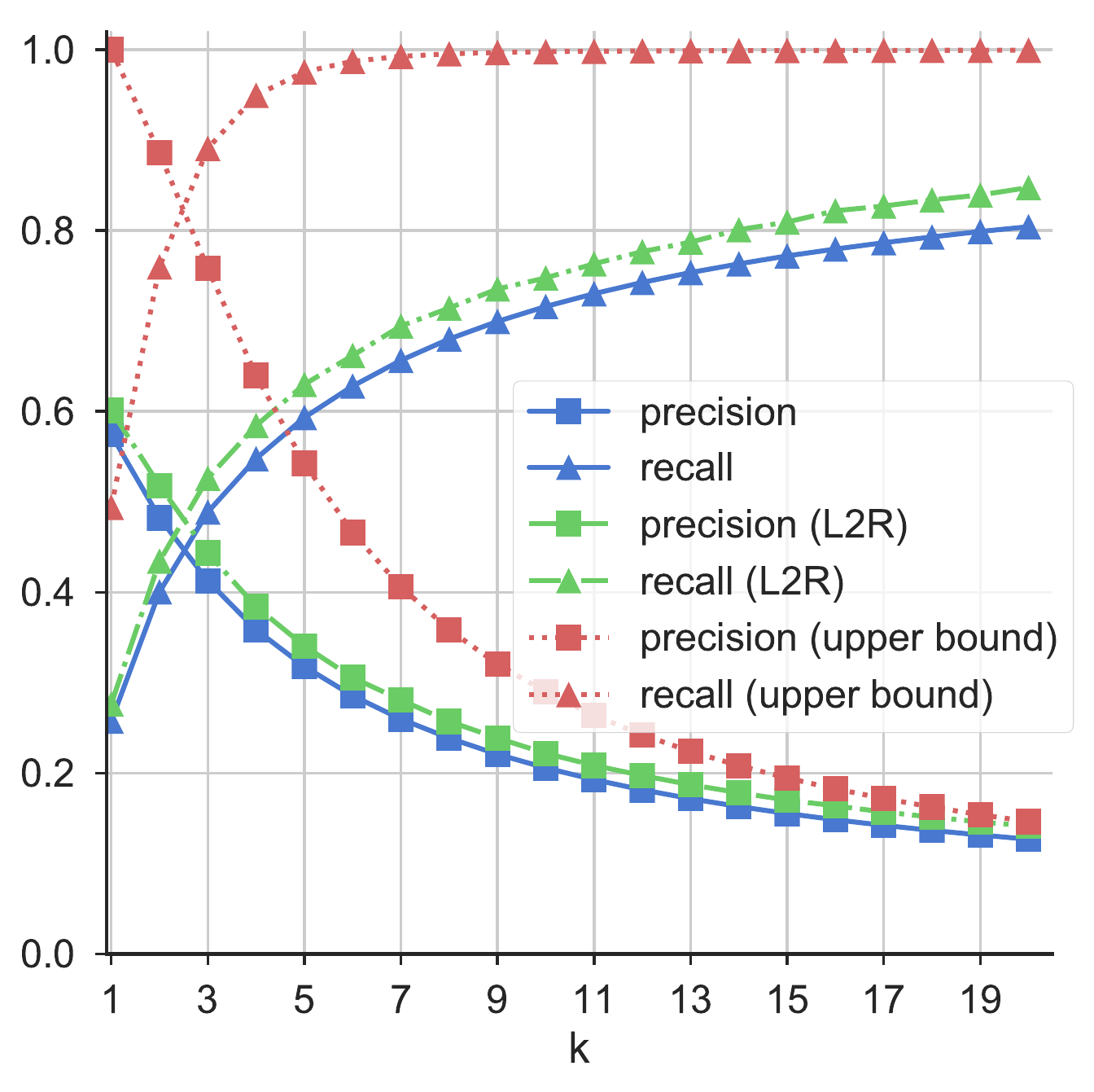}
	\label{fig:count_based}
}%
\hfill
\subfigure[Generalizing counts via CF (\Secref{sec:Generalizing via collaborative filtering})]{
	\includegraphics[width=0.55\columnwidth]{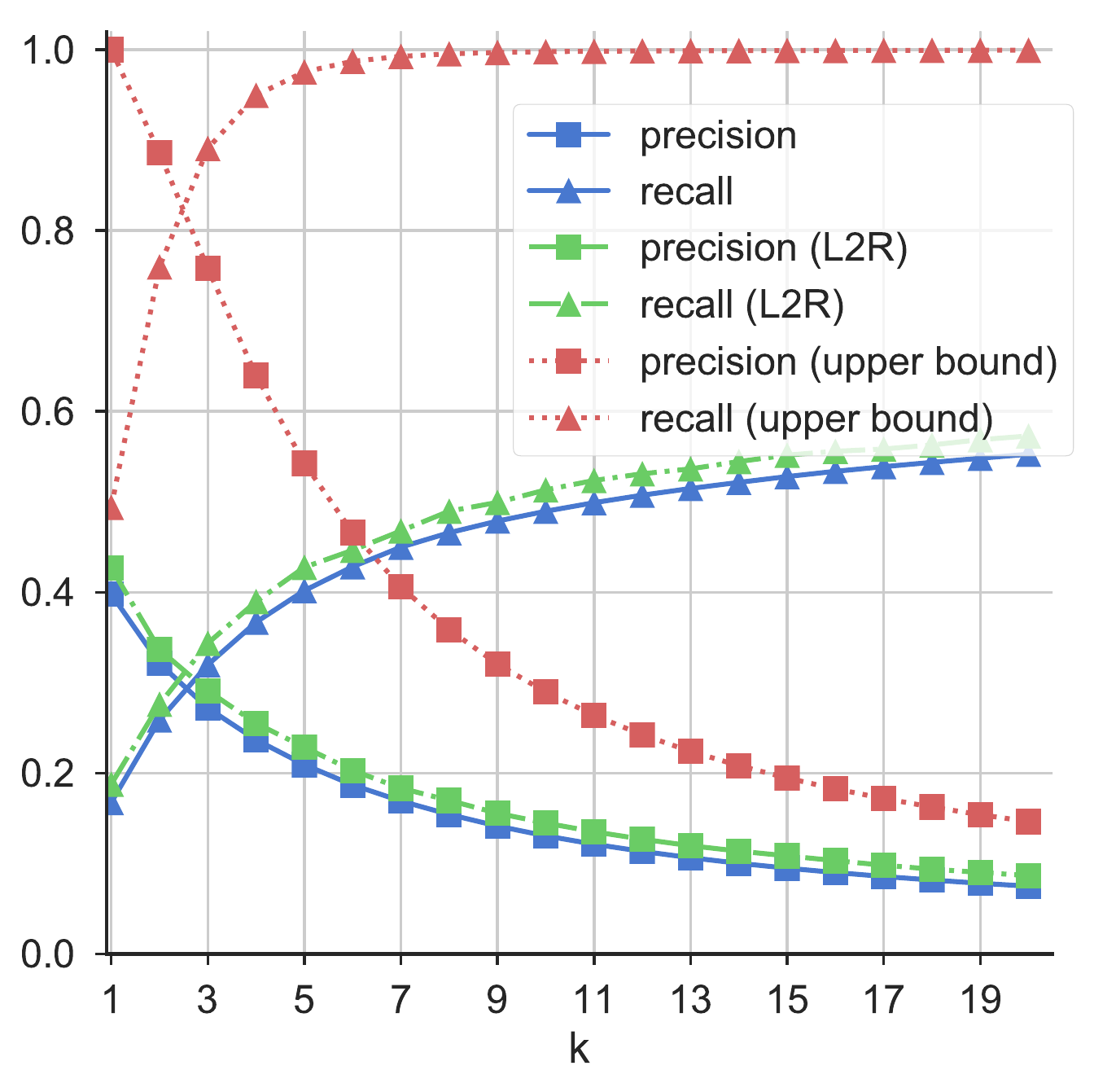}
    \label{fig:cf_based}
}%
\hfill
\subfigure[Human evaluation (\Secref{sec:Evaluation by human experts})]{
	\includegraphics[width=0.35\columnwidth]{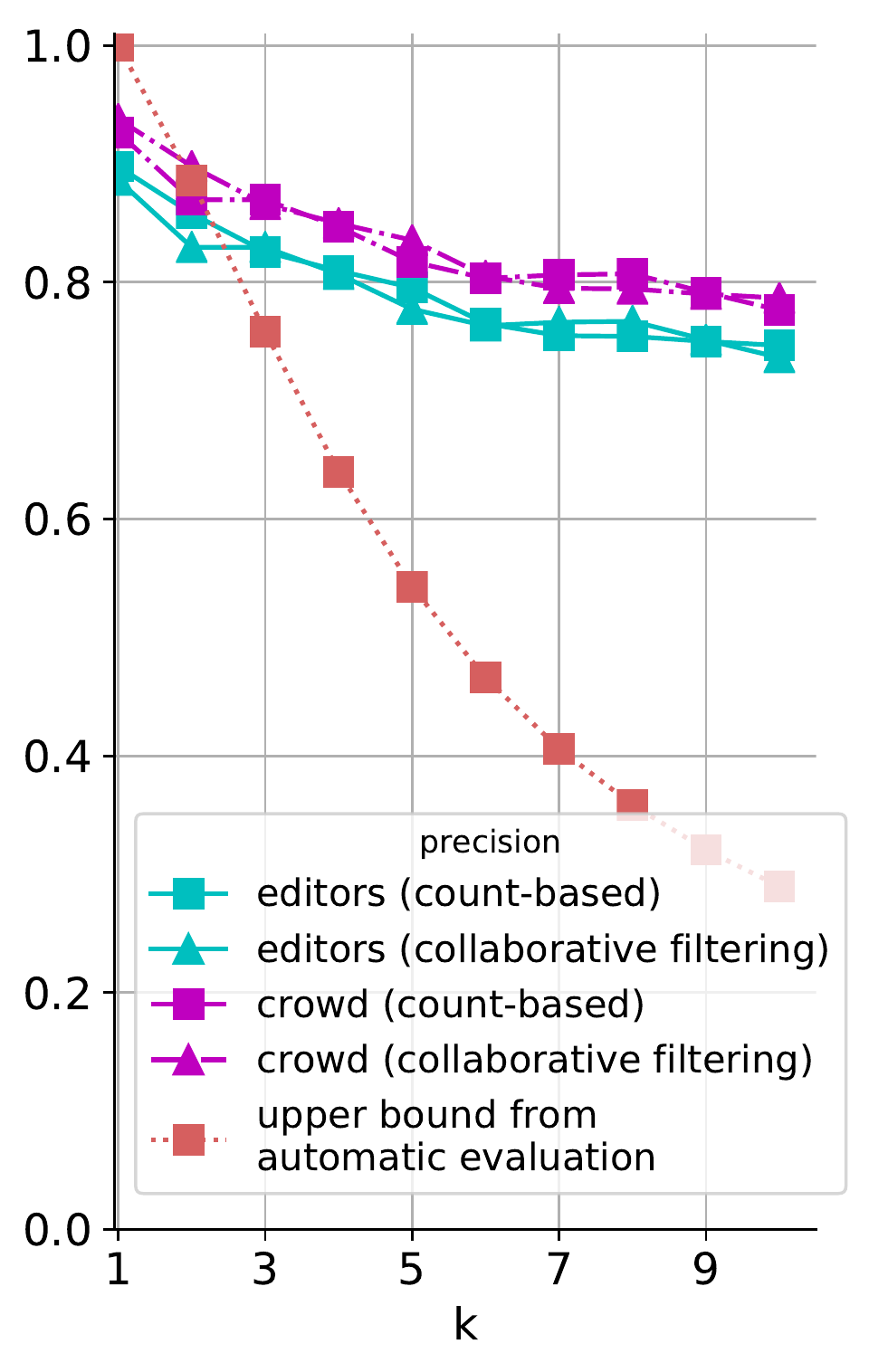}
    \label{fig:human_eval}
}
\vspace{-4mm}
\caption{Precision and recall as a function of the number of recommended sections $k$, for all methods we evaluate.}
\label{fig:precision_recall_curves}

\end{figure*}

While precision and recall trivially lie between 0 and 1, not the entire range is feasible:
if the number of sections in the testing article is $n$, then recall@$k$ is upper\hyp bounded by $k/n$, which is less than 1 if $k<n$ (\ie, if we are allowed to make only few recommendations);
and precision@$k$ is upper\hyp bounded by $n/k$, which is less than 1 if $n<k$ (\ie, if the testing article has only few sections).
When assessing our performance (\Figref{fig:precision_recall_curves}), it is important to keep these upper bounds in mind.

%%%%%%%%%%%%%%%%%%%%%%%%%%%%%%%%%%%%%%%%%%%%%%%%%%%%%%%%%%%%%%

\section{Evaluation: Article-based recommendation}
\label{sec:Evaluation: Article-based recommendation}
This section discusses the results obtained using the article\hyp based recommendation methods of \Secref{sec:Article-based recommendation}, which suggest sections based on the textual and section content of the input article.
%We designed mixed evaluation strategies to meet the requirements of the models described, and for each of them,
As we shall see, neither method yields results that would make it suitable for a real-world deployment scenario.

\begin{table*}
  \caption{Top 5 section recommendations for the Wikipedia article \cpt{Lausanne}, according to various methods.}
  \label{tab:examples}
  \begin{tabular}{llll}
    \toprule
    \textbf{Topic modeling} & \textbf{Article-based} & \textbf{Category--section} & \textbf{Generalizing counts via}\\
    (\Secref{sec:Using topic modeling}) & \textbf{collab.\ filtering} (\Secref{sec:Using collaborative filtering}) & \textbf{counts} (\Secref{sec:Using category--section counts}) & \textbf{collab.\ filtering} (\Secref{sec:Generalizing via collaborative filtering})\\
    \midrule
    \cpt{History} & \cpt{History of the document} & \cpt{History} & \cpt{History} \\
    \cpt{Sports} & \cpt{Famous resident} & \cpt{Demographics} & \cpt{Career} \\
    \cpt{Awards} & \cpt{Communes without arms} & \cpt{Economy} & \cpt{Personal Life} \\
    \cpt{Medal Summary} & \cpt{Content and importance} & \cpt{Education} & \cpt{Honours} \\
    \cpt{Statistics} & \cpt{Player movement} & \cpt{Politics} & \cpt{Career Statistics} \\
    \bottomrule
  \end{tabular}
\end{table*}

\subsection{Using topic modeling}
\label{sec:art-based eval: Using topic modeling}

\Figref{fig:baseline_topic} summarizes the performance achieved by the topic\hyp modeling approach (\Secref{sec:Using topic modeling}).
Precision and recall are plotted in blue, the theoretical upper bounds (\cf \Secref{sec:Evaluation dataset}) in red.
The method achieves a maximum recall (at $k=20$) of 30\%.
Precision is considerably lower, at 20\% for the single top recommendation, and it decreases to less than 5\% for $k=20$.

%Inspecting results, we found that the rankings produced by this model essentially retrieve secs that appear in many arts; in other words, this model degenerates into a plain popularity\hyp based recommender

Inspecting the recommended sections manually, we found that the major shortcoming of this method---rather intuitively---is that it can only capture textual, not taxonomic, similarity.
A typical example of how the topic\hyp based method may yield suboptimal results is given in \Figref{fig:overview} (top right): sections recommended for the input article \cpt{Stanford, California}, are effectively sourced from the article \cpt{Stanford University} (\eg, the section \cpt{Student life}), as these two articles are very similar in terms of words contained, and therefore in terms of the topics extracted via LDA.
In fact, however, the two articles call for widely different sections, one being a town, the other, a university.

\subsection{Using collaborative filtering}
\label{sec:art-based eval: Using collaborative filtering}

The article\hyp based collaborative\hyp filtering method (\Secref{sec:Using collaborative filtering}) makes it hard to generate recommendations for articles that were not part of the training set. For this reason, we designed an evaluation setup where the goal is to reconstruct half of the sections for the articles of the testing set. 
In the training phase, we included all the articles of the filtered subset (2.9M), and we removed 50\% of the sections only from the rows that are part of the testing set. 
% We excluded the articles of the testing set with less than two sections, so we have at least one element to learn from and one to test on. This step reduced the testing set to 277K (11.5\%) articles.
We excluded the articles of the testing set with less than two sections, so we have at least one element to learn from and one to test on. This step reduced the testing set to 277K articles.

As described in \Secref{sec:Using collaborative filtering}, we trained the model by factoring the article\hyp by\hyp section ``rating'' matrix with alternating least squares. We treat the problem as an explicit feedback setup in which the article gave a favorable ``rating'' to its sections.
%We selected the optimal training parameters by minimizing the Mean Square Error on a subset of 20\% of the elements that we masked from the matrix.  We achieved the minimal error using 50 latent factors, a regularization parameter $\lambda$ of 0.05.

Although the experiment is intentionally designed to make the task easier (leaving half of the sections in the training set and ignoring articles with only one section), the performance of the recommendation is unsatisfactory.
In particular, precision is always below 0.2\%, and recall at $k=20$ recommended titles stays below 1.5\%.
Although these values are significantly better than a random baseline, where precision is below 0.002\% for all $k$, and recall for $k=20$ is 0.008\%, this method is not suitable to be used in a real-world scenario.
It is well known that matrix factorization techniques perform poorly in the face of highly sparse data (a problem commonly referred to as ``cold start''~\cite{Koren:2009:MFT:1608565.1608614}). As such, considering that a Wikipedia article contains 3.48 sections on average, the low precision and recall figures for this method are not unexpected.

% \subsection{Discussion}
% \label{sec:art-based eval: Discussion}

% \todo{
% TODO: MICHELE
% }

% \begin{itemize}
% \item these methods fail because...
% \item LDA - [michele]
% \item Collaborative filtering: see sections distribution, not enough information, reference paper
% \end{itemize}

%%%%%%%%%%%%%%%%%%%%%%%%%%%%%%%%%%%%%%%%%%%%%%%%%%%%%%%%%%%%%%

\section{Evaluation: Category-based recommendation}
\label{sec:Evaluation: Category-based recommendation}

We now proceed to evaluating the category\hyp based methods of \Secref{sec:Category-based recommendation}.
As in the previous evaluation, we report the precision and recall for different values of $k$ to show how the system behaves with different lengths of the recommendations list.%
\footnote{Results available at \url{https://github.com/epfl-dlab/structuring-wikipedia-articles}}
%Given the distribution of the sections count in the articles (\Figref{fig:sections_distribution}), we include the upper limit both for precision and recall. This curve describes the ideal performance of the system for the scores at different values of $k$.

\subsection{Using category--section counts}
\label{sec:cat-based eval: Using category--section counts}

Using the training set and the cleaned category network, we compute the probability $P(S|C)$ of each section $S$ in each category $C$, as described in \Secref{sec:Using category--section counts}. This step generates scores for more than 191M category--section pairs, from which we extract a mapping between each category in Wikipedia and a set of sections sorted by their relevance. 

%Differently from the previous experiment, using the categories as an entry point, allows us to design a different test and evaluate the recommendations on all the sections' titles. Indeed, the performance metrics of this experiment are based on the reconstruction of all the sections present in the articles of the testing set.
Using an 80/15/5 train\slash test\slash validation split, we compute $P(S|C)$ for all $(S,C)$ based on the training set.
To make recommendations for a given article $A$, rather than category $C$, we combine the recommendations from all of $A$'s categories via a simple, unweighted sum of all the category\hyp specific scores.
% in two ways:
% (1)~via a simple, unweighted sum of the the scores from all categories, and
% (2)~via a weighted sum based on learning\hyp to\hyp rank (\Secref{sec:Combining recommendations from multiple categories}).

%In our evaluation, we generate the recommendations for an article by merging the relevant sections of it categories both with simple sum and with a weighted sum based on L2R as described before. 
% * <glederre@gmail.com> 2017-10-31T20:05:12.216Z:
% This is the first time that you talk about a weighted sum. Are you talking about the transformation of the score into the probability?
% ^.

\Figref{fig:count_based} shows that, with this method, precision for the unweighted sum reaches 57\% for the first recommended section; at $k=20$ recommendations, recall reaches 80\%.
We consider this performance sufficient for deploying the method in practice.

% - Since we are doing automatic evaluation, one sample is positive only if we have a complete match on the string. This 

% \begin{itemize}
% \item training setup, raw probability: big category are more noisy and contribute less
% \item merging [michele]
% \item results plot
% \end{itemize}

% \begin{figure}
% \includegraphics[width=0.8\columnwidth]{FIG/error}
% \caption{Mean Square Error with different values of $\lambda$, $\alpha$, and latent factors}
% \label{fig:error}
% \end{figure}

\subsection{Generalizing via collaborative filtering}
\label{sec:cat-based eval: Generalizing via collaborative filtering}

To model the relevance of a section in a category with collaborative filtering (CF), it is important to choose an adequate ``rating'' representation. 
We found that the best setup to training the model is to use the probabilities generated by the count\hyp  based approach while including only the top 100 sections for each category. Since we use matrix factorization to transfer information (and thus sometimes noise) between categories, the drawback is that this method is more sensitive to noise than the previous approach. Including only the top 100 sections per category helps to reduce the impact of large categories that describe very generic concepts and contain thousands of sections. After this step, we normalize rows to sum to 1 to make them comparable. %This scaling is necessary because the categories with many articles can have long tails of infrequent sections, which takes away probability mass from the very frequent items.

As in the case of article\hyp based recommendations, we factorize the matrix with alternating least squares, but for this problem we modeled the CF ratings (relevance of a section) as implicit feedback. Contrary to explicit feedback, where the ratings are represented as an explicit signal provided by the user, the implicit feedback is based on the presence or absence of an event (a section in this case).
Besides showing better results, in this case, the ratings are not an explicit representation of the sections appearing in a certain article, but rather an implicit signal of the most relevant sections for a specific category.
We selected the training parameters with the same approach described before (\Secref{sec:art-based eval: Using collaborative filtering}).
%, and %as shown in \Figref{fig:error} we minimized the mean squared error with 0.01 as regularization parameter $\lambda$, 0.5 for $\alpha$ and using 30 latent factors.

\Figref{fig:cf_based} shows that collaborative filtering on the category setup performs much better than the article-based approach, but, maybe unexpectedly, it does not outperform the recommendations generated by the counting method.
It seems that the aforementioned propagation of noise outweighs the advantage of generalization.

\subsection{Combining recommendations from multiple categories}
\label{sec:Eval: Combining recommendations from multiple categories}
\Figref{fig:count_based} and \Figref{fig:cf_based} showcase also the precision--recall curves (in green) for the learning\hyp to\hyp rank (L2R) merging strategy described in \Secref{sec:Combining recommendations from multiple categories}. For both the count\hyp based and the collaborative\hyp filtering method, L2R provides a performance boost both in terms of precision (3\%) and recall (4\%).
To interpret the results, we inspected the optimized feature weights and found the learned model assigns weights to the recommendation rankings that are inversely proportional to the size of a category, and directly proportional to the Gini coefficient (\ie, purity) of a category (\Secref{sec:Cleaning the category network}).

\subsection{Evaluation by human experts}
\label{sec:Evaluation by human experts}
The automatic evaluation described in the previous sections has the fundamental limitation that it introduces a negative bias in the results, thus deflating our performance numbers:
Many relevant sections are not yet present in the testing articles.
Also, as there are no strict rules followed by the Wikipedia editors when it comes to formulating section names, a lot of section occurrences are simple syntactical and semantical variations over the same concept (\eg, \cpt{Award, Awards, Prizes}). The automatic evaluation is based on the exact matches of the string representations, so it fails to return a positive result for both syntactic and semantic variations.
For these reasons, we also had humans evaluate our section recommendations, to assess the full performance of our models.
We run our tests with two groups: experienced Wikipedia editors ($N=11$) and crowd workers ($N=43$). Respectively, the two groups evaluated the section recommendations for 88 and 1,024 articles. The crowd workers of the second group were recruited on CrowdFlower, a popular crowdsourcing platform, and received \$0.10 per evaluated article.
To perform this evaluation, we developed a custom Web UI which shows both the recommended section and the Wikipedia article in the same browser window, allowing the evaluators to quickly iterate over the recommendations.
Each article in the evaluation sample was assigned the top 10 recommendations from our best performing methods: count-based and collaborative filtering with learning\hyp to\hyp rank merging (as described in
\Secref{sec:Combining recommendations from multiple categories}).
%As the recall is inherently bounded by the performance of our automatic methods, in this experiment
Computing recall would require the ground truth set of all relative sections for all test articles, which we do not have, so
we focus only on the precision of our recommendations as perceived by the human evaluators.

As evident from
\Figref{fig:human_eval},
precision@$k$ is substantially higher in the human, compared to the automatic, evaluation,
from the point of view of both expert editors and crowd workers, with a precision@1 of 89\% and 96\%, respectively, and precision@10 of 81\% and 72\%. The count-based and collaborative\hyp filtering methods are indistinguishable here, but there is a noticeable gap between the evaluations generated by the two groups. Such a difference is not surprising, as expert editors have more strict criteria when it comes to selecting sections for a certain article, while the crowd workers assessed the quality of the recommendations more intuitively based on their pertinence for the given article.

Finally, the drop in precision as $k$ grows is a sign that evaluators did not lazily provide all\hyp positive labels.

%\note{Random evaluation}

\section{Discussion and future work}
\label{sec:Discussion}

%\note{0.75p}

Wikipedia, though big, is notoriously incomplete, with fewer than 1\% of articles receiving a quality label of at least ``good'', and 37\% considered stubs \cite{Wikipedia:Statistics}. The encyclopedia needs a significant amount of contributions in order to expand its existing articles and to increase their quality.

%The human-centric approach of creating content on Wikipedia comes into play when building recommendation systems to help editors find what article to create next \cite{Wulczyn16, Cosley07}. 

In this research, we propose a methodology to help Wikipedia editors
%learn how to
expand an already existing article by recommending to them what sections to add to the articles they are editing.
In the rest of this section, we discuss how future work can address some of the challenges faced by this research, as well as opportunities for further research.

\xhdr{Entry point}
In this paper, we focused on one specific pathway for contributions, where users start from a Wikipedia article as an entry point.
Given our above\hyp described setup, we note, however, that we could just as well use categories as entry points; \eg, the user could specify that they want to start a new article in a certain category, and our method could suggest appropriate sections.
The methods of \Secref{sec:Category-based recommendation} can easily be adapted to this use case, as they already rely on the categories the input article belongs to.

\xhdr{Improving section recommendations} There are several ways in which section recommendations can be improved in practice: 
%\vspace{-3mm}
\begin{itemize}
\item Semantically related sections are considered as independent sections in the current approach. The quality of our section recommendations could be improved by grouping these related sections (\eg, \cpt{Works} and \cpt{Bibliography}; or \cpt{Life} and \cpt{Early life}) together and recommending only one of them.
\item Providing more information to the editor, beyond the section title, can help the editor learn what is expected of a section. For instance, information about the average section length, the pages a section usually links to, \etc, can be helpful information for the editor.
%\note{Sourced from external languages. Sections Alignment}

\item Our method currently falls short in specifying the order in which the recommended sections should be added. The order of sections in an article can, however, be important, especially for longer Wikipedia articles with many sections.
\item The recommendations in this research are built using the category network information within a given Wikipedia language. With more than 160 actively edited Wikipedia languages, there is also a wealth of information in other  languages that can help when expanding articles in a given language. In the future, it will be key to consider the multilingual aspect of Wikipedia and benefit from it, especially as these recommendations should be applicable to medium\hyp size and smaller Wikipedia languages, where the category network may not be fully developed or the amount of content is simply too limited to source useful recommendations from within the same language. 
\item The current count-based recommendation approach works well on frequently occurring sections. However, one of Wikipedia's strengths is the unique view of editors reflected in its content. Research on understanding how the long tail of less frequent sections can be mined to recommend infrequent but important sections will be critical in improving the recommendations.
\end{itemize}

\xhdr{Human-centered recommendation approach} Wikipedia is a human\hyp centered project, where the editors' judgment, deliberation, and curiosity play key roles in how the project is shaped and content is created. This human-centric approach creates an enormous opportunity for automatic recommendations in the sense that striving for perfect precision does not need to become the center-piece of the research, while high recall will allow us to not miss out on the most important recommendations that we would miss otherwise. The same human\hyp centered approach to content creation, however, demands interpretability of recommendations, as any edit on Wikipedia may be challenged and reasoning behind why a specific piece of content is added to the project is key in empowering the editors to use such recommendations. The count-based recommendation system developed in this research is a good starting point for allowing us to provide interpretable reasons for our recommendations.

\xhdr{Beyond section recommendations}
Building a recommendation system for expanding a Wikipedia article will not end at recommending what sections can be added to the article. Past research has developed technology for adding hyperlinks to Wikipedia articles \cite{Paranjape16}, and further research can result in more comprehensive recommendations telling editors what images, citations, external links, \etc, to add as well.

\xhdr{Production system}
Finally, our system will be truly useful only once it is incorporated at scale in a production system.
We believe our system's performance numbers are sufficiently high for deployment, with a precision@10 of around 80\% (\Figref{fig:human_eval}), so we aim to incorporate our recommender system into Wikipedia's Visual Editor in the near future.

\section{Conclusion}
\label{sec:Conclusion}

In the present paper, we have introduced the task of recommending sections for Wikipedia articles.
Sections are the basic building blocks of articles and are crucial for structuring content in ways that make it easy to digest for humans.
We have explored several methods, some that are based on features derived immediately from the input article that is to be enriched with sections (\eg, content and pre-existing sections), and others that instead generate recommendations by leveraging Wikipedia's category system.
Our evaluation clearly shows that the category\hyp based approach is superior, reaching high performance numbers in an evaluation by human raters (\eg, precision@10 around 80\%).
We hope to deploy our system \textit{in vivo} in the near future, in order to contribute to the growth and maintenance of one of the greatest collaborative resources created to date.

\section*{Acknowledgments}
We thank Diego S\'aez-Trumper from the Wikimedia Foundation, as well as J\'er\'emie Rappaz and Gael Lederrey from EPFL, for thoughtful discussions.
We are also grateful to the Wikimedia editor community members who helped with the human evaluation.
% This research has been supported in part by XX, YY, and ZZ.

\bibliographystyle{ACM-Reference-Format}
\bibliography{bibliography} 

\end{document}